\documentclass[aps, pra, showpacs,twocolumn,superscriptaddress]{revtex4-1}

\usepackage{epsfig}
\usepackage{graphicx}
\usepackage{dcolumn}
\usepackage{color}
\usepackage{comment}
\usepackage[normalem]{ulem}
\usepackage{breakcites}
\usepackage[hyperindex,breaklinks]{hyperref}
\graphicspath{{./figures/}}

\begin{document}

\title{Superconducting Acousto-optic Phase Modulator}

\author{Ayato Okada}
\affiliation{Research Center for Advanced Science and Technology~(RCAST), The University of Tokyo, 4-6-1 Komaba, Meguro-ku,
	Tokyo 153-8904, Japan}

\author{Rekishu Yamazaki}
\email[]{rekishu@icu.ac.jp}
\altaffiliation[Current address: ]{College of Natural Sciences, International Christian University, Mitaka, Tokyo 180-8585, Japan}
\affiliation{Research Center for Advanced Science and Technology~(RCAST), The University of Tokyo, 4-6-1 Komaba, Meguro-ku,
	Tokyo 153-8904, Japan}
\affiliation{PRESTO, Japan Science and Technology Agency, Kawaguchi-shi, Saitama 332-0012, Japan}

\author{Maria Fuwa}
\affiliation{Research Center for Advanced Science and Technology~(RCAST), The University of Tokyo, 4-6-1 Komaba, Meguro-ku,
	Tokyo 153-8904, Japan}

\author{Atsushi Noguchi}
\altaffiliation[Current address: ]{Komaba Institute for Science (KIS), The University of Tokyo, Meguro-ku, Tokyo 153-8902, Japan}
\affiliation{Research Center for Advanced Science and Technology~(RCAST), The University of Tokyo, 4-6-1 Komaba, Meguro-ku,
	Tokyo 153-8904, Japan}
\affiliation{PRESTO, Japan Science and Technology Agency, Kawaguchi-shi, Saitama 332-0012, Japan}

\author{Yuya Yamaguchi}
\affiliation{Network System Research Institute, National Institute of Information and Communications Technology,
	4-2-1, Nukuikitamachi, Koganei, Tokyo 184-8795, Japan}

\author{Atsushi Kanno}
\affiliation{Network System Research Institute, National Institute of Information and Communications Technology,
	4-2-1, Nukuikitamachi, Koganei, Tokyo 184-8795, Japan}

\author{Naokatsu Yamamoto}
\affiliation{Network System Research Institute, National Institute of Information and Communications Technology,
	4-2-1, Nukuikitamachi, Koganei, Tokyo 184-8795, Japan}

\author{Yuji Hishida}
\affiliation{Advanced ICT Research Institute, National Institute of Information and Communications Technology,
	588-2 Iwaoka, Nishi-ku, Kobe, Hyogo 651-2492, Japan}

\author{Hirotaka Terai}
\affiliation{Advanced ICT Research Institute, National Institute of Information and Communications Technology,
	588-2 Iwaoka, Nishi-ku, Kobe, Hyogo 651-2492, Japan}

\author{Yutaka Tabuchi}
\affiliation{Research Center for Advanced Science and Technology~(RCAST), The University of Tokyo, 4-6-1 Komaba, Meguro-ku,
	Tokyo 153-8904, Japan}

\author{Koji Usami}
\affiliation{Research Center for Advanced Science and Technology~(RCAST), The University of Tokyo, 4-6-1 Komaba, Meguro-ku,
	Tokyo 153-8904, Japan}

\author{Yasunobu Nakamura}
\affiliation{Research Center for Advanced Science and Technology~(RCAST), The University of Tokyo, 4-6-1 Komaba, Meguro-ku,
	Tokyo 153-8904, Japan}
\affiliation{RIKEN Center for Quantum Computing (RQC), Wako, Saitama 351-0198, Japan}

\date{\today}

\begin{abstract}
    We report the development of a superconducting acousto-optic phase modulator fabricated on a lithium niobate substrate.  A titanium-diffused optical waveguide is placed in a surface acoustic wave resonator, where the electrodes for  mirrors and an interdigitated transducer are made of a superconducting niobium titanium nitride thin film.  The device performance is evaluated as a substitute for the current electro-optic modulators, with the same fiber coupling scheme and comparable device size.  Operating the device at a cryogenic temperature ($T=8$~K), we observe the length--half-wave-voltage~(length--$V_\pi$) product of 1.78~V$\cdot$cm.  Numerical simulation is conducted to reproduce and extrapolate the performance of the device.  An optical cavity with mirror coating on the input/output facets of the optical waveguide is tested for further enhancement of the modulation efficiency.  A simple extension of the current device is estimated to achieve an efficient modulation with $V_\pi=$~0.27~V.
\end{abstract} 


\maketitle

\section{Introduction}
The advancement of optical communication technology using optical fibers is the core of the modern ultra-high-speed networks. One of the optical devices which play a central role in the optical communication is the optical modulator that converts an electrical signal into an optical signal. A wide range of electo-optic modulators~(EOMs) including thin-film dielectrics~\cite{stenger2013, wang2018}, semiconductors~\cite{ogiso2017, dong2012}, electro-optic polymer in slot waveguides~\cite{palmer2014}, and electro-optic polymer in plasmonic waveguide~\cite{heni2016} have been developed to further push its bandwidth as well as the energy efficiency.   For a conventional traveling-wave EOM, the half-wave voltage ($V_\pi$), which is defined as the necessary voltage applied to shift the optical phase by a half of the wavelength, is known as one of the indices characterizing the performance of the optical modulator.  The half-wave voltage  can be lowered in principle by extending the interaction length between the light and the applied microwave signal. However, in practice, the microwave signal is attenuated by the propagation loss of the electrodes in the device chip, which limits the ultimate device length and the half-wave voltage~\cite{Doi2006}. Moreover, the electrical bandwidth of the modulator is hindered by the profound electrical loss at higher frequencies, making a trade-off between the efficiency and bandwidth.  In general, a longer device has a smaller half-wave voltage, while a shorter device has a wider bandwidth.  It is important to mitigate the electrical loss for a highly-efficient wide-bandwidth device.

As a measure to overcome this limitation, superconducting electrodes and the operation at low temperature can be employed to mitigate the ohmic loss. Several earlier studies have been conducted for the exploitation of a high-$T_c$ superconducting film~\cite{yoshida1993, rozan1999}. However, optical modulators using superconductors have not yet been put to a commercial use.  The number of optical channels for the mobile connections are still increasing due to the rapid progress of the internet-of-things (IoT) technology, and it becomes more and more important to be able to reduce the energy consumption in electrical and optical communication devices.   As in the case of single-flux-quantum logic circuits for computation~\cite{Holmes2013}, it could be an effective solution to use superconducting circuits even with the cooling overhead.  

The use of acoustic waves, more precisely the photoelastic effect, for the optical modulation has long been overlooked, and there has been tremendous progress recently. The acoustic waves typically have much lower velocity than that of the electromagnetic waves, and their wavelength at the GHz frequency range approximately matches that of the optical waves.  The short wavelength enables tight confinement of the acoustic mode, resulting in a significant modulation with small device size. Also, the piezo-acoustic devices have a long history with matured technologies and well-studied materials~\cite{Datta1986, Campbell1998}.  Various useful designs of the piezoelectric transducers, acoustic mirrors, and acoustic resonators are readily available. Many acoustic materials are also known to have a high internal quality factor (Q-factor), reaching above a billion~\cite{Galliou2013}. These well-developed tools and knowledge in piezo-acoustic devices should accelerate the development of novel acousto-optical devices.

Integrated acousto-optical devices on lithium niobate~(LN) are of particular importance, and various developments, including the acousto-optic frequency shifter~\cite{Shao2020, Yu2021}, and the acousto-optic modulators~\cite{Cai2019, Sarabalis2020} are reported.  Many of these devices also utilize the thin-film LN on various insulators to further confine the acoustic mode. Other recent activities on the integrated acousto-optical devices are compiled in Ref.~\cite{zhu2021}

Additionally, the development of coherent conversion between microwave and optical photons are recently on call, stimulated by the rapid progress in the field of quantum computing~\cite{arute2019} and its extension to the quantum network~\cite{awschalom2019}. A superconducting resonator coupled to an electro-optic microresonator is theoretically investigated~\cite{javerzac2016}, and experimentally demonstrated~\cite{witmer2017, fan2018, Mckenna2020, Holzgrafe2020}.  Highly efficient acousto-optical devices at room temperature~\cite{Shao2019} and in cryogenic environment with superconducting electrodes~\cite{Jiang2020} are reported.

In this paper, we develop an acousto-optic phase modulator consisting of superconducting electrodes and a titanium-diffused (Ti-diffused) optical waveguide on a LN substrate to reduce the half-wave voltage of the modulator.  The optical waveguide is placed in a surface-acoustic-wave (SAW) resonator and the photoelastic effect induced by the SAW is employed as a modulating mechanism, along with the electro-optic (Pockels) effect utilized in the conventional EOM. By using the SAW whose energy is localized on the substrate surface, it is possible to efficiently modulate the optical signal guided in the waveguide. The Ti-diffused waveguide also provides an opportunity for excellent connectivity with fiber optics.   The length--half-wave-voltage (length--V$_\pi$) product obtained is more than 10-fold smaller than that of the commercially available traveling-wave EOM, with the device size smaller by about fifty times. We also develop a facet-coated optical waveguide cavity with the finesse of $\mathcal{F}=43$, which could additionally reduces the half-wave voltage approximately by $2\mathcal{F}/\pi$.  

\begin{figure*}[tbp!]
	\includegraphics[width=13.2cm]{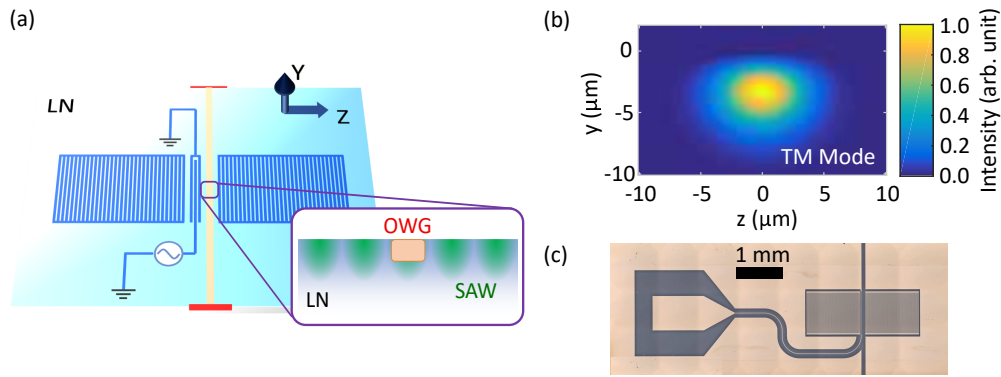}\\
	\caption{(a) Schematic of the superconducting acousto-optic phase modulator architecture.  A surface-acoustic-wave (SAW) resonator is fabricated on a Y-cut lithium niobate~(LN) substrate with the SAW propagation direction in the $Z$-direction. Both the interdigitated-transducer (IDT) and mirror electrodes~(blue) are made of a superconducting niobium titanium nitride~(NbTiN) film of 50-nm thick.   A titanium-diffused (Ti-diffused) waveguide~(orange) is fabricated on the same substrate with the optical propagation in the $X$-direction.  End surfaces of the substrate on both ends of the waveguide are polished to optical quality.  Additional mirror coating~(red) can be applied to form a waveguide optical cavity. The zoomed section shows a schematic of the cross section of the sample across the optical waveguide.  The optical waveguide~(OWG) is designed to be placed at the anti-node of the $u_z$-component of the SAW standing waves in the SAW resonator.  (b) Simulated optical intensity distribution in the Ti-diffused waveguide for the transverse-magnetic~(TM) mode.  The substrate surface is located at $y=0~\mu$m.  We estimate the mode diameter of $2w_0= 6.7~\mu$m and $9.7~\mu$m for the $y$- and $z$-directions, respectively. (c)  Optical micrograph of a sample chip. Light-colored part on the chip is the NbTiN film. The optical waveguide is located on the vertical gap between the Bragg mirrors of the SAW resonator.   }\label{fig1}
\end{figure*}

\section{Superconducting acousto-optic phase modulator: device description}
A schematic of the superconducting acousto-optic phase modulator is shown in Fig.~\ref{fig1}(a).  In order to realize highly efficient optical modulation by elastic waves, a system composed of a SAW resonator and an optical waveguide, is fabricated on a Y-cut LN substrate.  LN has a wide transparent window in the optical domain along with large piezoelectric and photoelastic coefficients.  The SAW resonator is fabricated with the SAW propagation direction in the $Z$-axis of the LN crystal.  We choose the device coordinate system ($x$, $y$, and $z$) to coincide with the crystal axes ($X$, $Y$, and $Z$) from here on.  Both the interdigitated transducer~(IDT) and Bragg-mirror electrodes are made of superconducting niobium titanium nitride~(NbTiN) film. The NbTiN film has a superconducting transition at the temperature of around 12~K for the thickness of 50~nm, which is higher than the operating temperature of our refrigerator, typically running at 8~K (Montana Instruments). The Ti-diffused waveguide is also fabricated on the same substrate with the optical propagation direction along the $X$-axis of the LN crystal. The Rayleigh-type SAW used in the experiment has the displacement field components denoted as $u_i$ $(i=y,z)$ and the electric potential $\phi$. The optical waveguide is patterned in between the SAW mirrors and designed to sit at the anti-node of the $u_z$-component of the SAW standing wave.  

When a radio-frequency (RF) signal, tuned to the resonance frequency of the SAW resonator, is applied to the IDT, standing waves of the SAW are excited in the SAW resonator.  An optical signal is input from one of the end facets of the waveguide.  The deformation and strain due to the SAW cause the modulation of the refractive index of the optical waveguide, resulting in the phase modulation of the input optical signal by the SAW frequency.   The modulation can be enhanced by the optical cavity, which effectively increases the interaction length between the SAW and optical input signal.

\section{Experiment}
\subsection{Device fabrication and sample preparation}
First, an optical waveguide is fabricated on a LN substrate using the Ti diffusion method. The waveguide is nearly a single mode for both the transverse-electric~(TE) and transverse-magnetic~(TM) polarizations at the laser wavelength of $\lambda_{\mathrm{opt}}=1064$~nm.  The optical field distribution of the TM mode, obtained by numerical simulation (OptiBPM), is shown in Fig.~\ref{fig1}(b). We obtain the fundamental mode $1/e^2$ diameter (defined at $1/e^2$ of the maximum power density) of $2w_0= 6.7~\mu$m and $9.7~\mu$m, for the $y$- and $z$-directions, respectively. 

Next, the SAW resonator is fabricated with a 50-nm-thick superconducting NbTiN thin film. The resonator is designed for the SAW wavelength $\lambda_{\mathrm{SAW}}=40$~$\mu$m, corresponding to the resonance frequency of approximately 87~MHz.  The SAW wavelength is chosen so that the half-wave of the SAW is large enough compared to the optical mode diameter in the waveguide.  After the SAW resonator fabrication, the substrate is diced and the end facets of the optical waveguide are polished to optical quality. Figure~\ref{fig1}(c) shows an optical micrograph of a typical device. 

\begin{figure*}[tbp!]
	\includegraphics[width=13.3cm]{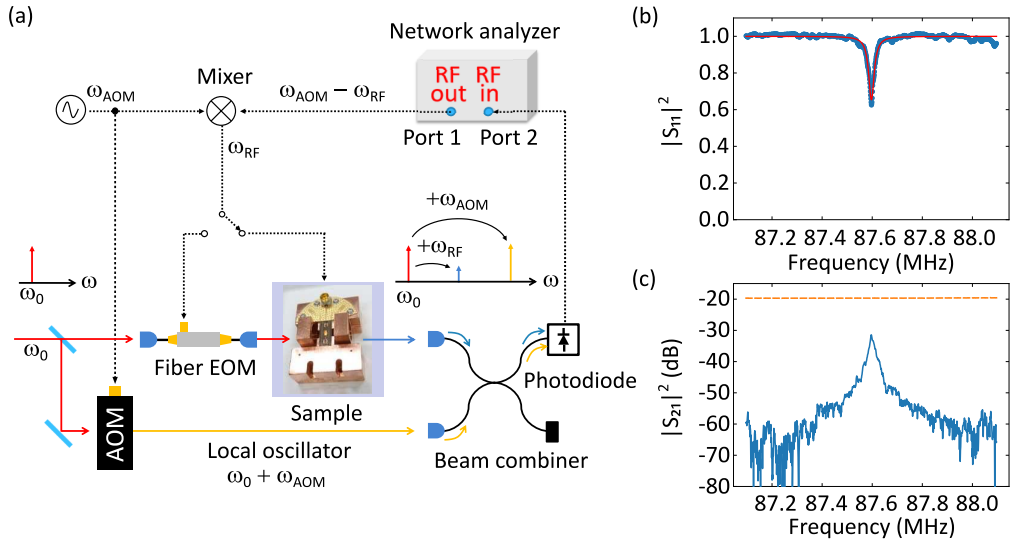}\\
	\caption{(a) Experimental setup: As shown in the inset picture, the device chip is fixed on a sample holder with a SMP connector and mounted inside a refrigerator.  For the RF spectroscopy, the sample is directly connected to port~1 of the network analyzer for the reflection measurement.  
	For the optical spectroscopy, an RF signal from port~1 of the network analyzer is mixed with another signal from a function generator to drive either the SAW-based modulator or a commercial EOM.  The optical signal is input to the modulator and the output signal is combined with an optical local oscillator before the measurement with a photodiode. The inset frequency diagram indicates the output signal and LO frequencies arriving at the photodiode. The photodiode signal is sent to port~2 of the network analyzer to determine the phase modulation imparted by the modulators. (b) Reflection spectrum of the SAW resonator obtained from RF spectroscopy. The blue dots are data and the red solid line is a Lorentzian fit to the data, from which we obtain the total mechanical loss rate of the SAW resonator $\Gamma/2\pi = 26.4$~kHz and the SAW-resonator quality factor of 3,300.  (c) Sideband signal ($S_{21}$) obtained in the heterodyne measurement.  The blue solid and orange dashed lines correspond to the heterodyne signal obtained for the SAW device and the commercial EOM, respectively. While the EOM shows almost no frequency dependence, the SAW device shows clear signal enhancement at the SAW resonance frequency.   Difference between the two signals at the SAW resonance frequency is 11.8~dB.  }\label{fig2}
\end{figure*}

Before performing RF and optical measurements in the cryogenic environment, the device is glued to a sample holder, as shown in the inset picture of Fig.~\ref{fig2}(a), and fixed to the cold finger of the refrigerator.  The free-space laser light is coupled into and out of the optical waveguide through collimating lenses mounted on the sample holder.   More details on the sample fabrication and preparation are compiled in Appendix~A.

\subsection{RF spectroscopy of the SAW resonator}
RF reflection of the SAW resonator is measured using a network analyzer.  The sample is directly connected to port~1 of the network analyzer for the reflection measurement.  The loss due to the ohmic resistance of the NbTiN electrodes is large at room temperature, and the signal from the SAW resonator could not be found. As the sample temperature was lowered, a dip in the RF reflection spectrum appeared around 11~K, as expected from the superconducting transition of NbTiN.  At temperature below 11~K, the shape of the reflection spectrum remains almost unchanged. The RF reflection spectrum of the SAW resonator measured at 7.8~K is shown in Fig.~\ref{fig2}(b). From the fitting, we obtain the resonance frequency $\Omega/2\pi = 87.6$~MHz and the Q-factor of 3,300, with the internal dissipation rate $\Gamma_\mathrm{in}/ 2\pi=23.9$~kHz, and external coupling rate $\Gamma_\mathrm{ex}/2\pi=2.5$~kHz. 

\subsection{Optical spectroscopy and measurement of $V_{\pi}$}
The SAW excitation by the RF signal input results in the phase modulation of the optical signal in the waveguide. We first evaluate the magnitude of the phase modulation by a heterodyne measurement~\cite{okada2018}. In order to quantify the depth of the phase modulation by the SAW, we perform the same heterodyne measurement on a reference EOM (Photline NIR-MPX-LN-10) and compare the sideband signal strength.  The half-wave voltage of the reference EOM is pre-determined by measuring the sideband and carrier amplitudes using a scanning Fabry-P\'{e}rot resonator. 

The optical setup used for the measurement is shown in Fig.~\ref{fig2}(a). The input optical signal polarization is in the $y$-direction to excite the fundamental TM mode of the optical waveguide.  The light emitted from the optical waveguide is coupled to a single-mode fiber, mixed with a local-oscillator~(LO) light by a beam combiner, and detected by a photodiode. The LO light is prepared by partially splitting the incident light (frequency $\omega_0$) to the optical waveguide and shifting the frequency with an acousto-optic modulator by  $\omega_{\mathrm{AOM}}$.  As shown in the inset frequency diagram of Fig.~2(a), the first blue sideband generated by the phase modulation creates a beat signal with the LO beam at frequency $\omega_\mathrm{AOM}-\omega_\mathrm{
RF}$.  The signal is appropriately filtered and measured with the network analyzer. 

The blue solid and orange dashed lines in Fig.~\ref{fig2}(c) show the results of the sideband generation for the SAW modulator and the commercial EOM, respectively, while sweeping the excitation RF frequency with a network analyzer. For the SAW modulator, the sideband amplitudes are maximized at the SAW resonance frequency, while the EOM showing a flat spectrum within the frequency range of the data. Both measurements are taken with the same RF power. At the SAW resonance, the sideband power difference of 11.8~dB is measured. The half-wave voltage of the EOM is $V_{\pi}=4.8$~V at the driving frequency of 87~MHz. Since the sideband power is inversely proportional to the square of the half-wave voltage, the half-wave voltage of the SAW modulator for the single-pass is determined as $V_{\pi}= 18.7$~V. Due to a large difference in the device length, the product of the interaction length and half-wave voltage, which is one of the figure of merit for optical modulators, can be calculated for a better comparison. For the reference EOM, whose pigtail to pigtail packaging size is 100~mm, we estimate a device length of 50~mm, which is a typical length for these devices.  We calculate length--$V_{\pi}$ product of 24~V$\cdot$cm.  For the SAW modulator, the interaction length of the SAW and light is assumed to be equal to the Bragg-mirror width $W=$~950~$\mu$m, which gives the length--$V_{\pi}$ product of 1.78~V$\cdot$cm.  The small length--$V_{\pi}$ product shows the efficient modulation delivered by the SAW modulator,  achieved by  the high Q-factor of the SAW resonator with superconducting electrodes.  We discuss further enhancement of the modulation efficiency with a use of  an optical cavity later.  
 
\begin{figure*}[tbp!]
	\includegraphics[width=13.2cm]{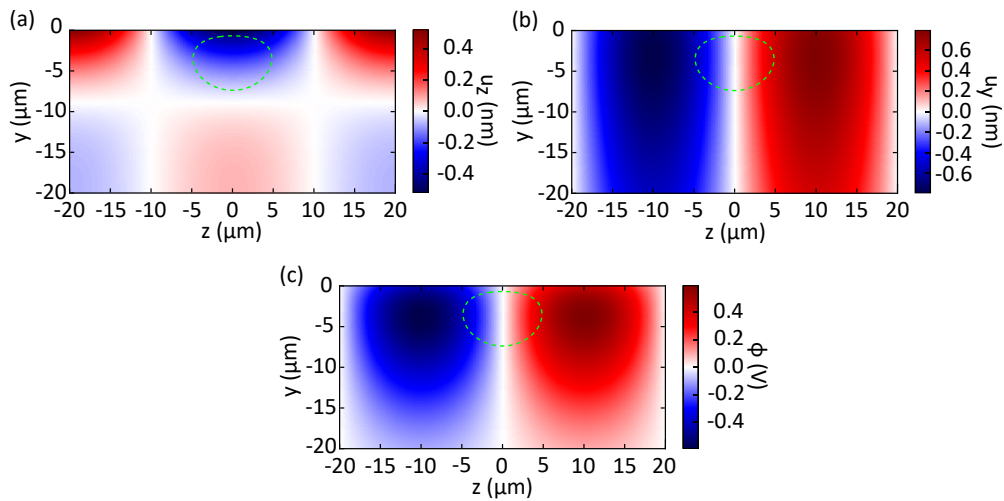}\\
	\caption{Calculated displacement and electric potential in the SAW resonator. The color scale represents the displacement and the electric potential of the SAW when resonantly driven by an external RF signal generator with an amplitude of 1~V. (a) Longitudinal displacement $u_z$, (b) transverse displacement $u_y$ and (c) electric potential $\phi$. The green dashed lines represent the mode spread (defined at $1/e^2$ of the maximum power density) of the light propagating in the Ti-diffused optical waveguide. }\label{fig3}
\end{figure*}

\section{Theoretical analysis}
In order to quantitatively understand the observed phase modulation, we examine the spatial distribution of the SAW-induced phase modulation imparted on the light propagating in the optical waveguide. 

The refractive index variation due to the photoelastic and electro-optic effects in terms of the electric field $E_k= \frac{\partial \phi}{\partial x_{k}}$  and strain tensor $S_{kl} = \frac{1}{2} \left( \frac{\partial u_{k}}{\partial x_{l}} + \frac{\partial u_{l}}{\partial x_{k}}\right)$ reads,
\begin{equation}
\Delta\!\left(\frac{1}{n^2}\right)_{ij}=r_{ijk}E_k + p_{ijkl}S_{kl},
\end{equation}
where $r_{ijk}$ is the electro-optic tensor at zero strain (clamped) and $p_{ijkl}$ the photoelastic tensor at constant electric field~\cite{jazbinsek2002}.  Here, we use the coordinate indices~(${i,j,k,l}={1,2,3}$) instead of (${x,y,z}$). For the case where the TM mode ($y$-polarization) is chosen for the input and output optical modes, the refractive index variation in terms of the SAW displacement $u_i$ and electric potential $\phi$ is,  
\begin{equation}\label{dn}
\delta n_y=\frac{n_y^3}{2}\left[r_{22k}\frac{\partial \phi}{\partial x_k} -\frac{ p_{22kl}}{2}\left(\frac{\partial u_k}{\partial x_l}+\frac{\partial u_l}{\partial x_k}\right)\right],
\end{equation}
where $n_y$ is the unperturbed refractive index for the $y$-polarization.  

\begin{table}[tbp!]
\centering
\caption{Material constants of LiNbO$_3$ used for the determination of SAW-induced refractive index~\cite{jazbinsek2002}. }
\begin{tabular}{cc}
\hline
Constant & Value\\
\hline
$n_y$ & 2.23\\
		$p_{2222}$ & $-0.0261$\\
		$p_{2233}$ & 0.0832\\
		$p_{2223}$ & 0.1335\\
		$r_{222}$ & $3.40\times10^{-12}$~m/V\\
	    $r_{223}$ & $9.10\times10^{-12}$~m/V \\
\hline
\end{tabular}
  \label{table1}
\end{table}

The contributions of the terms proportional to $u_i$ and $\phi$, have different relative phases.  Numerical results of the distributions of longitudinal displacement $u_z$($=u_3$),  transverse displacement $u_y$($=u_2$), and electric potential $\phi$ in the vicinity of the optical waveguide are shown in Fig.~\ref{fig3}. The dashed lines in Fig.~\ref{fig3} indicate the spread of the optical mode propagating in the optical waveguide.  From the variation of the refractive index $\delta n_y$ calculated using Eq.~\ref{dn} for a given RF input power, the half-wave voltage $V_{\pi}$ is calculated from the following formula,
\begin{equation}
\pi = \delta n_y (V_{\pi})\,k_{\mathrm{opt}} W, 
\end{equation}
where $k_{\mathrm{opt}}$ is the optical wavenumber.   
Material constants used in the calculation are listed in Table~\ref{table1}~\cite{jazbinsek2002}.  

The contribution of the electro-optic effect, driven by the piezoelectrically-induced electric field, and that of the photoelastic effect from the strain are found to be comparable. We calculate $V_{\pi}= 17.3$~V from the numerical simulation, slightly lower than the observed value.  In the same simulation, we also found that a misplacement of the optical waveguide by $-0.75$~$\mu$m in the $z$-direction results in a diminished modulation with $V_\pi=$~18.7~V.  We estimate the misalignment between the optical waveguide and the SAW resonator during the fabrication to be at most $1~\mu$m, and the numerical simulation is consistent with our measurement. 

As shown in Fig.~\ref{fig3}, $u_z$ has a mirror symmetric distribution around $z=0$, while $u_y$ and $\phi$ show asymmetric distributions.  Different strain components ($S_{kl}$) and electric fields ($E_k$) derived from the displacement and electric potential also have either symmetric or asymmetric distribution around $z=0$.  Among them, only the ones with symmetric distribution contributes to the phase modulation.  We find $S_{23}$ and $E_3$ to be the main contributors for the photoelastic and electro-optic phase modulations, respectively.  We also find that the center of the optical mode is approximately 4~$\mu$m below the surface, while the largest variation of the refractive index is found nearly at the surface. Possible improvement may be found by lifting the optical mode location towards the surface by constructing a ridge-structured waveguide. 


\begin{figure*}[tbp]
	\centering\includegraphics[width=8.4cm]{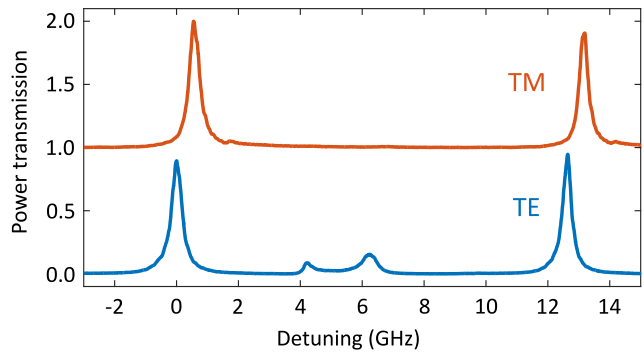}\\
	\caption{Optical waveguide cavity transmission spectra for the TE and TM modes shown in blue and red lines, respectively.  The detuning refers to the first peak of the TE mode.  The amplitude of the transmission spectra is normalized with the maximum peak height of the TM mode. The TM mode signal is offset by unity.  The input laser frequency is swept more than the free-spectral-range of the cavity.  The cavity finesse $\mathcal{F}=43$ and the linewidth $\kappa/2\pi=284$~MHz are obtained for the TM mode.  The quality factor of the cavity is $Q = 1.0\times10^6$. We observe nearly the same performance for the two input optical polarizations.  The higher spatial modes are observed in the TE mode presumably due to a smaller mode area in the optical waveguide.}  \label{fig4}
\end{figure*}


 \section{Optical waveguide cavity}
 In order to  further improve the performance of the modulator, one can apply mirror coating on both end facets of the optical waveguide to form an optical cavity. The ratio of the output optical power in the first sideband for a device with the optical cavity ($P_{\mathrm{SB, cavity}}$) to that for a device with only the waveguide ($P_{\mathrm{SB, waveguide}}$) can be derived as,
 \begin{equation}
 \frac{P_{\mathrm{SB, cavity}}}{P_{\mathrm{SB, waveguide}}}=\left(\frac{2\mathcal{F}}{\pi}\right)^2\frac{\kappa_\mathrm{ex}^2}{\Omega^2+(\kappa/2)^2},
 \end{equation}
 where $\kappa_{\mathrm{ex}}$, $\kappa$, and $\mathcal{F}$ are the external coupling rate, total dissipation rate, and finesse of the optical cavity, respectively.  
When the SAW frequency $\Omega$ is less than the linewidth of the cavity $\kappa$~$(\Omega < \kappa)$, the power enhancement of approximately $\left(\frac{2\mathcal{F}}{\pi}\right)^2$, equivalently $V_\pi$ reduction of $2\mathcal{F}/\pi$, can be achieved at the critical coupling condition ($\kappa_\mathrm{ex}=\kappa/2$).  Contrary, when the SAW frequency is larger than $\kappa$, the enhancement drops as the density of states for the scattered photon diminishes rapidly outside the cavity linewidth.  
 
 To show this possibility, we apply dielectric multi-layer film coating with reflectivity $R=95~\%$ on both end facets of the optical waveguide chip used for the current device.  Transmission spectra of the optical waveguide cavity for the TE and TM modes are shown in Fig.~\ref{fig4}.   We obtain the free-spectral ranges of 12.6 and 12.1~GHz for the TE and TM modes, respectively, which is consistent with the birefringence of LN.  For the TM mode, we observe the finesse of $\mathcal{F}=43$ ($Q = 1.0\times10^{6}$), with the corresponding total cavity dissipation rate $\kappa/2\pi=284$~MHz, and find a negligibly small difference of the finesse between the two input laser polarizations.  The cavity finesse did not seem to change from the room temperature to  the cryogenic temperature during the cool-down, either.  The propagation loss of the optical waveguide is estimated to be less than $0.2$~dB/cm.  The worst case would limit the maximum finesse of $\mathcal{F}\approx 150$ ($Q = 1.4\times10^{6}$), for a 5-mm-long cavity at the critical coupling.  The effect of the optical cavity can be understood as a simple elongation of the interaction length by approximately $2\mathcal{F/\pi}$, as long as the SAW frequency $\Omega$ is much smaller than the cavity linewidth.

 \section{Discussion} 
The modulation efficiency $V_{\pi}=$ 18.7~V for the current device is partially limited by the short SAW cavity width~(950~$\mu$m) along the optical waveguide.  By extending the length of the SAW cavity to 50~mm, which is about a typical length in commercial EOMs, we expect to obtain $V_{\pi}=$ 2.58~V for a single-pass device.  Additional optical cavity implementation  with the finesse of $\mathcal{F}=15$ ($Q = 1.4\times10^{6}$), corresponding to the channel propagation loss of 0.2~dB/cm, would lead to $V_{\pi}=$ 0.27~V.  For the future mobile transport networks, a narrow-band and high-efficient optical modulator is one of the key components for the radio-over-fiber~(RoF) links~\cite{Murata2020}. In those links, narrow-band radio signals are transmitted through optical fibers by using electro-optic~(EO) and opto-electric~(OE) conversion without frequency mixing in electrical domain. The resonance-type narrow-band modulator with high efficiency, presented here, could potentially be applied to such future communication networks.

While we routinely observe the Q-factor of an acoustic resonator exceeding $10^5$ with aluminum electrodes on a bulk LN in a cryogenic environment, the Q-factor of the current device is limited to 3,300 even with the use of NbTiN superconducting electrodes.  We presume the degradation of the Q-factor originates from the reactive ion etching of NbTiN electrodes on LN with CF$_4$ plasma during the fabrication.  We expect the Q-factor to reach the same level as other acoustic resonators made with alternative fabrication techniques, enhancing the modulation efficiency. The  high Q-factor acoustic resonator in a cryogenic environment may also pave the way towards using these devices in the quantum regime.

There are a number of reports, similarly improving the efficiency of the optical modulation. Among different types of EOMs, devices made of thin-film LN has attracted much attention recently. Development of various EOMs including LN ridge-structured waveguides on silicon \cite{Stenger2019, Wang2018b},  and Ti-diffused waveguides \cite{Yamaguchi2019} are reported with estimated $V_{\pi} = $~2.5, 1.4, and 1.2~V, respectively.  With an optical resonator, an estimated $V_{\pi} = $ 11~V is also reported \cite{wang2018}.  All of these measurements were conducted at room temperature.  One of the drawbacks of the current device may be a relatively large optical mode width of the Ti-diffused waveguide, limiting the modulation efficiency. The large extent of the optical mode could be an obstacle for increasing the acoustic frequency.  Incorporating this thin-film LN technology, 
at the expense of the fiber connectivity, may be the key to improve the performance of the current device further.
   
In another research area, where the focus is on coherent wavelength conversion between optical and microwave photons, tremendous efforts have been made to utilize both electro-optic and photoelastic effects along with superconducting circuits to efficiently convert the photons while mitigating the loss.  SAW resonators coupled with an optical racetrack resonator \cite{Holzgrafe2020}, optical and microwave hybrid-ring resonator \cite{fan2018}, and photonic-phononic crystal resonator \cite{Jiang2020}, have been reported with estimated $V_{\pi} = $~ 0.088, 0.032, and 0.024~V, respectively.  Many of these works are at the frontier of efficient optical modulation. However, the use of low-temperature superconducting electrodes requiring a temperature below 4~K and the intricacy of fabrication could be overwhelming and may not be the near-future solution for the replacement  of the commercial EOMs.

While the current device shows moderate modulation efficiency, it requires much simpler fabrication.  The facet mirror coating shown here can also be implemented to various devices.  The well-developed fiber-based butting technique for efficient coupling into a Ti-diffused waveguide is available with or without the waveguide facet mirror coating to form an optical cavity.  The thin-film LN technology often suffers from the fiber connectivity due to the tight confinement of the optical mode and the size mismatch with standard single-mode fibers.  While the mode conversion is possible, it still remains a challenge to reach near unity efficiency.  These advantages make the current device attractive for a mass production and as a replacement of the commercial EOMs. 

\section{Summary}
In summary, we developed a superconducting acousto-optic phase modulator.  Operated at the cryogenic temperature of 8~K, the device showed a length--$V_{\pi}$ product of 1.78~V$\cdot$cm, which is more than 10-fold improvement compared to commercially available electro-optic modulators. Implementing an optical cavity in the current device structure is shown to further improve the overall performance.    

\begin{acknowledgments}
The authors acknowledge K. Kusuyama and K. Nittoh for the help in sample fabrication. This work was partly supported by JSPS KAKENHI (Grant No.~26220601), NICT NEXT-project, JST PRESTO (Grant No.~JPMJPR1429, JPMJPR1667), and JST ERATO (Grant No.~JPMJER1601).
\end{acknowledgments}

\appendix
\renewcommand{\thefigure}{S\arabic{figure}} 
\setcounter{figure}{0}
\renewcommand{\theequation}{S\arabic{equation}} 
\setcounter{equation}{0}
\renewcommand{\thetable}{S\arabic{table}} 
\setcounter{table}{0}

\maketitle

\section{Fabrication and sample preparation}\label{appendix1}

An optical waveguide is fabricated using the titanitum diffusion method, in which we sputter a Ti strip with a width of 3~$\mu$m and a thickness of 90~nm on a LN substrate and diffuse the titanium at the temperature of 960$^\circ$C for 21 hours. 

Next, a 50-nm-thick superconducting NbTiN thin film is sputtered on the entire surface of the substrate on which the optical waveguide is fabricated. The SAW resonator is patterned by electron beam lithography and dry-etched with CF$_4$. The resonator is designed for the SAW wavelength $\lambda_{\mathrm{SAW}}=40$~$\mu$m.  The SAW wavelength is chosen, so that the half-wavelength of the SAW is large enough compared to the optical mode diameter in the waveguide.  Using the SAW velocity of 3,488~m/s in LN, the corresponding resonance frequency is approximately 87~MHz. The interdigitated transducer and Bragg mirrors are 950~$\mu$m wide and have four~(two pairs) and two-hundred electrodes, respectively. The distance between the mirrors, measured between the closest electrodes, is 380~$\mu$m.  The angular misalignment between the optical waveguide and the SAW resonator is approximately within 0.06~degrees, which may shift the location of the SAW standing wave by 1~$\mu$m with respect to the optical waveguide. 

After the SAW resonator fabrication, the substrate is diced with a sintered diamond blade and the end facets of the optical waveguide are polished to optical quality. The length of the optical waveguide after polishing is 5.5~mm.  The waveguide is nearly a single mode for both the TE and TM polarizations at the laser wavelength of $\lambda_{\mathrm{opt}}=1064$~nm. 

The optical-quality polishing and mirror coating of the end facets of the optical waveguide are conducted externally (LAMBDA PRECISION CORPORATION).  The multi-layer mirror-coating consists of SiO$_2$, Al$_2$O$_3$, and Nb$_2$O$_5$ layers with a total thickness of 1830~nm.

Before performing RF and optical measurements in the cryogenic environment, the device is fixed to the measurement jig made of copper.  The sample is glued on the jig with varnish next to an external printed circuit board with a SMP connector soldered on it.  The RF input port for the IDT is wire-bonded to the external printed circuit board. Next, the jig is fixed to the cold finger of the refrigerator, and an alignment laser is introduced by butt-ending a bare fiber to the output of the optical waveguide.  An input collimator lens mounted on the side of the sample is adjusted to collimate the alignment beam out of the optical waveguide.  We then introduce a free-space laser to couple the beam into the waveguide through the input collimator and optimize the coupling by aligning the input laser to the alignment laser.  Once the input laser is optimized, the alignment laser is removed, and an output collimator lens is placed and adjusted.  After the optical alignment is completed, the sample is covered with a radiation shield, and the refrigerator is closed for cool-down. The free-space laser light is coupled into and out of the waveguide via the optical windows of the refrigerator throughout the experiment.

\section{Estimation of the SAW amplitude}
  In order to calculate the SAW amplitude in the SAW resonator, we use the relationship between the input RF power and the phonon number in the cavity.  First, the zero-point fluctuation amplitude of the SAW is defined as, 
  \begin{equation}
  U_{\mathrm{zpf}}=\sqrt{\frac{\hbar}{2\rho V_{\mathrm{mode}}\Omega}},
  \end{equation} 
  where  $\rho$, $V_{\mathrm{mode}}$, $\Omega$ are the mass density of the SAW substrate, the mode volume of the SAW resonator, and the SAW frequency, respectively.   The mode volume is defined as $V_{\mathrm{mode}}=\lambda L W$, where $\lambda$ is the SAW wavelength.  $L$ and $W$ are the length and width of the SAW resonator, respectively~\cite{schuetzs2015}.

  The amplitude of the SAW, $U_0$, is related to the number of phonons in the SAW resonator, $N$, by
  \begin{equation}
  U_0^2 =NU_{\mathrm{zpf}}^2 = \frac{N\hbar}{2\rho V_{\mathrm{mode}}\Omega}.
  \end{equation}

  The number of phonons in the single-port SAW resonator for the input power $P$ is given as,
  \begin{equation}
  N = \frac{\Gamma_{\mathrm{ex}}}{\Delta^2+(\Gamma^2/4)}\frac{P}{\hbar\Omega},
  \end{equation}
  where $\Delta = \omega_\mathrm{RF}-\Omega$ is the detuning between the RF input frequency $\omega_\mathrm{RF}$ and the SAW resonance frequency $\Omega$.  $\Gamma_\mathrm{ex}$ and $\Gamma$ are the external coupling and total loss rates of the SAW resonator, respectively.   
  At the SAW resonance frequency ($\Delta =0$),
  \begin{equation}
  N = \frac{4\Gamma_{\mathrm{ex}}}{\Gamma^2}\frac{P}{\hbar\Omega},
  \end{equation}
  and the SAW amplitude $U_0$ can be rewritten in a form,
  \begin{equation}
  U_0^2 = \frac{4\Gamma_{\mathrm{ex}}}{\Gamma^2}U_{\mathrm{zpf}}^2\frac{P}{\hbar\Omega}.
  \end{equation}

In the simulation, we use $U_0$ determined from the input power $P$ as the amplitude of the SAW. 


%


\end{document}